\documentclass[preprint,aps]{revtex4-1}

\usepackage{graphicx}
\usepackage{dcolumn}
\usepackage{bm}

\begin{document}

\title{Tailoring strain in SrTiO$_3$ compound by low energy He$^{{\rm +}}$ irradiation}

\author{S. Autier-Laurent, P. Lecoeur} 
\affiliation{Institut d'Electronique Fondamentale, IEF, UMR CNRS 8622, Univ. Paris Sud, Bat. 220, F-91405 Orsay Cedex, France} 
\author{O. Plantevin}\email{plantevin@csnsm.in2p3.fr}
\author{B. Decamps}
\author{A. Gentils}
\author{C. Bachelet} 
\author{O. Kaitasov} 
\affiliation{Centre de Spectrom\'etrie Nucl\'eaire et de Spectrom\'etrie de Masse, Universit\'e Paris-Sud, UMR CNRS-IN2P3 8609, 91405 Orsay Cedex France}
\author{G. Baldinozzi}
\affiliation{SPMS, UMR CNRS 8580, Ecole Centrale Paris, F-92295 Ch$\hat{a}$tenay-Malabry, France}

\date{\today}

\begin{abstract}
The ability to generate a change of the lattice parameter in a near-surface layer of a controllable thickness by ion implantation of strontium titanate is reported here using low energy He$^{{\rm +}}$ ions. The induced strain follows a distribution within a typical near-surface layer of 200 nm as obtained from structural analysis. Due to clamping effect from the underlying layer, only perpendicular expansion is observed. Maximum distortions up to 5-7\% are obtained with no evidence of amorphisation at fluences of 10$^{16}$ He$^{{\rm +}}$ ions/cm$^2$ and  ion energies in the range 10-30 keV.                    
               {\bf Accepted for publication in Europhysics Letter http://iopscience.iop.org/0295-5075}
\end{abstract}


\maketitle

\section{INTRODUCTION}
Induced strain in compounds derived from the perovskite structure is known to be an efficient parameter to tune their physical properties. Superconductors and manganites are very sensitive to such changes and drastic evolutions can be observed in their properties. Strain can be applied in different ways: chemical pressure by substituting cations in the structure, external strain applied with an anvil cell, or biaxial strain when deposited as thin film by hetero-epitaxy. In this last case, the strain due to the mismatch between the lattice parameter of the unclamped film and the substrate parameter can be used to permanently stabilize structures under pressure\cite{Prellier_JAP2001}. The epitaxial strain has even been shown to induce room-temperature ferroelectricity in strontium titanate, which is not normally ferroelectric at any temperature \cite{Haeni}. So far, for that purpose, only a limited series of substrates is able to fulfil the required conditions: to succeed, the lattice mismatch cannot be too large, to avoid extended default formation. It would be of great interest to find a way to tune the lattice parameter. The present paper reports on the investigation of the correlation of the lattice parameter change of He$^{{\rm +}}$ irradiated SrTiO$_3$ single crystals with the applied fluence and helium ion energy. Mechanism responsible for the observed strain distribution is discussed from structural analysis compared with simulations of the implanted profile. Within that frame, the strain distribution has to be considered as a consequence of defects introduction in the material. As will be shown, helium ion irradiation introduces mainly point defects such as vacancies which can lead to a completely new electronic structure. From that point of view, this method is much different from epitaxial strain.

\section{EXPERIMENTAL}
SrTiO$_3$ (STO) is, at room temperature, a cubic perovskite (Pm$\bar{3}$m) with a cell parameter of 3.905 $\AA$. This is one of the most used substrates for thin oxide film growth. The irradiated STO were 5$\times$5$\times$0.5 mm$^3$, (001) and (110) oriented crystals.
The samples were irradiated at CSNSM using the IRMA implantor \cite{Chaumont_NIMB1981}. For the (001) samples, a He$^{{\rm +}}$ beam of varying ion energy between 10 keV and 30 keV energy was used with an ion fluence of 10$^{16}$ He$^{{\rm +}}$ ions/cm$^2$. Different ion fluences between 10$^{15}$ He$^{{\rm +}}$ ions/cm$^2$ and 5$\times$10$^{16}$ He$^{{\rm +}}$ ions/cm$^2$ were used at an ion energy of 30 keV. The (110) oriented samples were irradiated with an energy of 30 keV and two ion fluences of 5$\times$10$^{15}$ He$^{{\rm +}}$ ions/cm$^2$ and 5$\times$10$^{16}$ He$^{{\rm +}}$ ions/cm$^2$. The implantations were performed with an ion current density below 3  $\mu$A/cm$^2$, at room temperature and under an angle of incidence of 7$^\circ$ to avoid canalization effects.
One (110) oriented STO sample was studied after helium ion implantation at a fluence of 5$\times$10$^{15}$ He$^{{\rm +}}$ ions/cm$^2$  using cross-sectional Transmission Electron Microscopy (TEM FEI Tecnai G$^{{\rm 2}}$ 20, 200 kV). It was only mechanically polished for that purpose after the observation that Ar$^{{\rm +}}$ ion polishing interferes with the observation by causing the formation of small cavities in the thin sample.
XRD experiments were performed using an X'PERT PANalytical Philips diffractometer equiped with a Cu anode (K$_\alpha$ radiation (8 keV)) in a $\theta$-2$\theta$ geometry close to the (002) Bragg peak for the (001) oriented crystals, and around the (220) Bragg peak for the (110) oriented sample. In plane parameter was deduced from asymetric reflection (202) in the case of STO(001) samples.

\section{RESULTS}
\subsection{SIMULATION}

\begin{figure}
\includegraphics[width=5in]{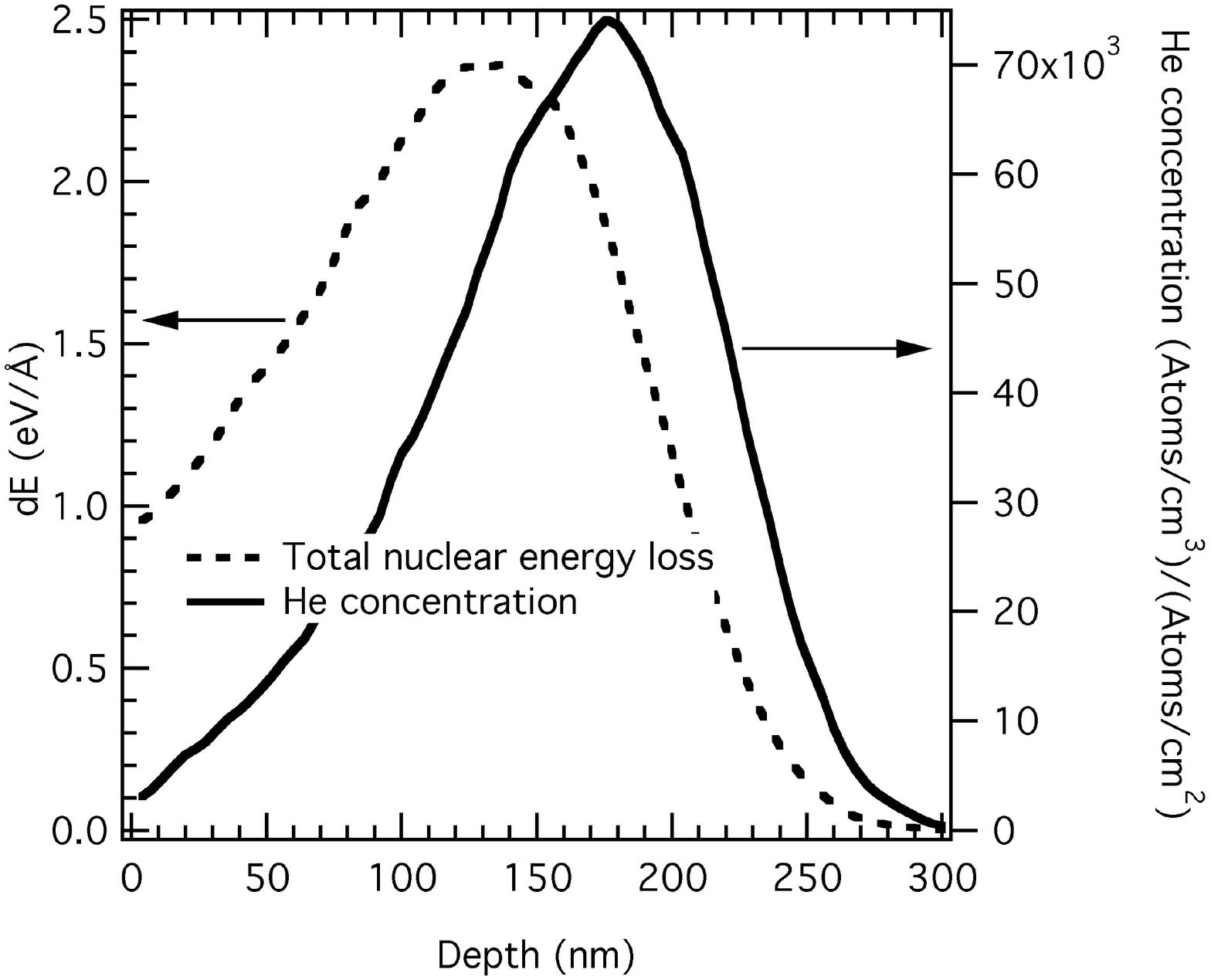}
\caption{SRIM simulation results of the nuclear energy loss and helium concentration in SrTiO$_3$ irradiated with 30 keV He$^{{\rm +}}$ ions}
\label{Fig1}
 \end{figure}

According to the SRIM simulation program\cite{SRIM} (http://www.srim.org) , the mean projected range for the helium ions at 30 keV, R$_p$, and straggling, $\delta$R$_p$, are 157 nm and 55 nm respectively. The simulation results are independent of crystalline orientation as the sample is treated as a mean medium without atomic structure. The maximum He concentration for an ion fluence of 10$^{16}$ He$^{{\rm +}}$ ions/cm$^2$ is 7.4$\times$10$^{20}$cm$^-$$^3$ at a depth of 176 nm, while the total vacancy concentration following the simulation is estimated to have a maximum of 2.4$\times$10$^{22}$cm$^-$$^3$ at a depth of about 130-140 nm, which corresponds to 50\% of the oxygen sites. This value is expected to be much lower due to direct vacancy-interstitial recombination  and only about 20\% of the defects is estimated to survive at room temperature as was assumed for example in MgO under similar irradiation conditions\cite{Schut}. This assumption leads to approximately 10\% of oxygen vacancy sites.  An experimental method which could be used to determine the formation of vacancy defects is positron annihilation spectroscopy. It has already be shown that cation-oxygen related vacancies were induced by Ar ion implantation in SrTiO$_3$  single crystals \cite{Gentils}. Both the profiles of He concentration and nuclear energy loss are shown in Fig.\ref{Fig1}. Nuclear energy loss is directly responsible for defect creation in the irradiated samples. From SRIM, the concentration of oxygen vacancies is estimated to be about twice the concentration of Sr and Ti vacancies (which are approximately equal), and the maximum in oxygen vacancy concentration is shifted slightly deeper (150 nm) due to a lower mobility edge. When the implantation energy is increased from 10 keV to 30 keV, one finds that the maximum of the nuclear energy loss is translated from 47 nm to 140 nm, while the FWHM of the profile increases from 75 nm to 160 nm.

\subsection{X-RAY DIFFRACTION}  
  
\begin{figure}
\includegraphics[width=8in]{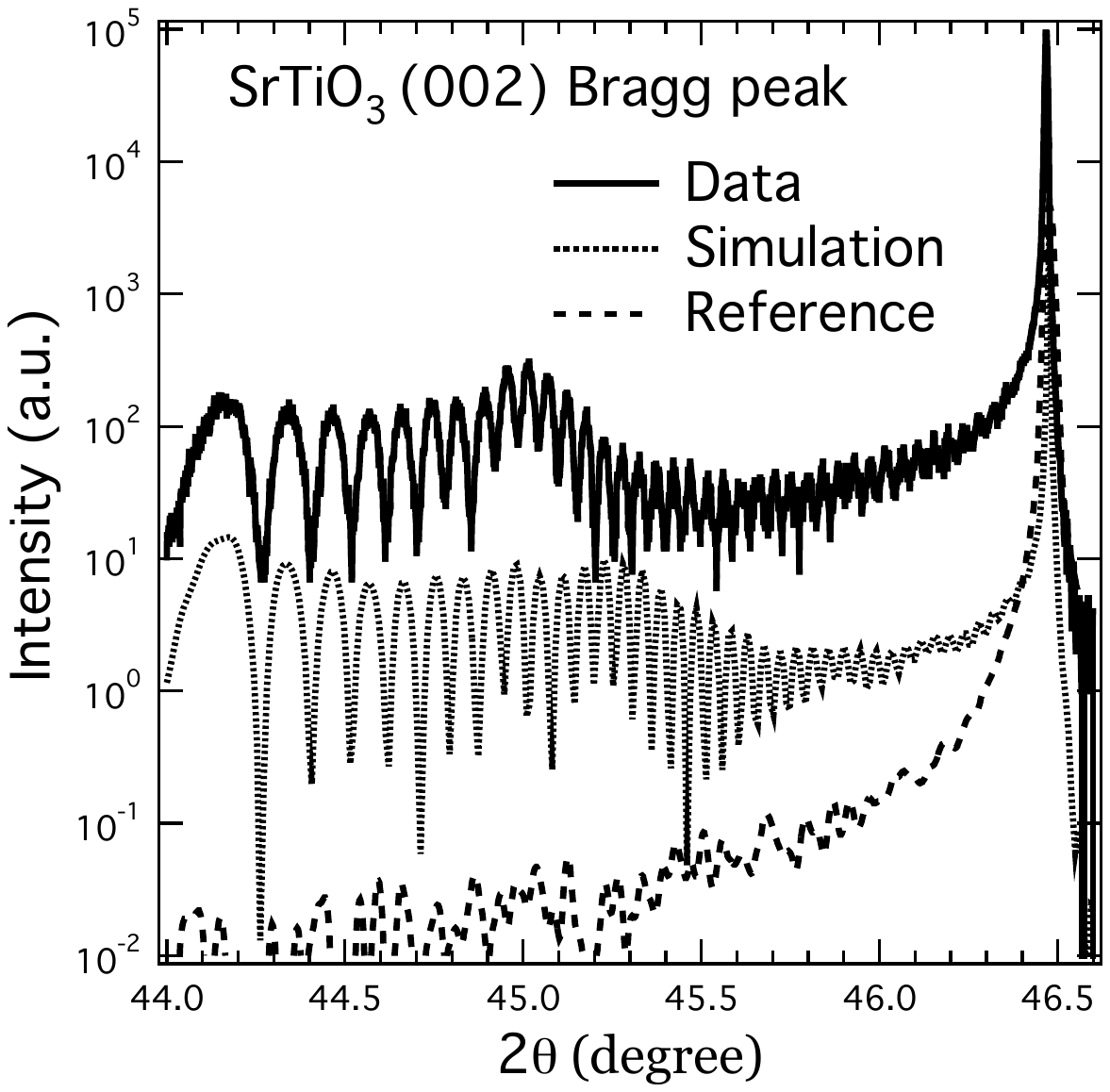}
\caption{Measured and simulated X-ray diffraction spectra of a (001) oriented SrTiO$_3$  sample irradiated with 10$^{16}$ He$^{{\rm +}}$ ions/cm$^2$ ions at 30 keV compared with a reference (non irradiated) sample.}
\label{Fig3}
 \end{figure}
 
\begin{figure}
\includegraphics[width=5in]{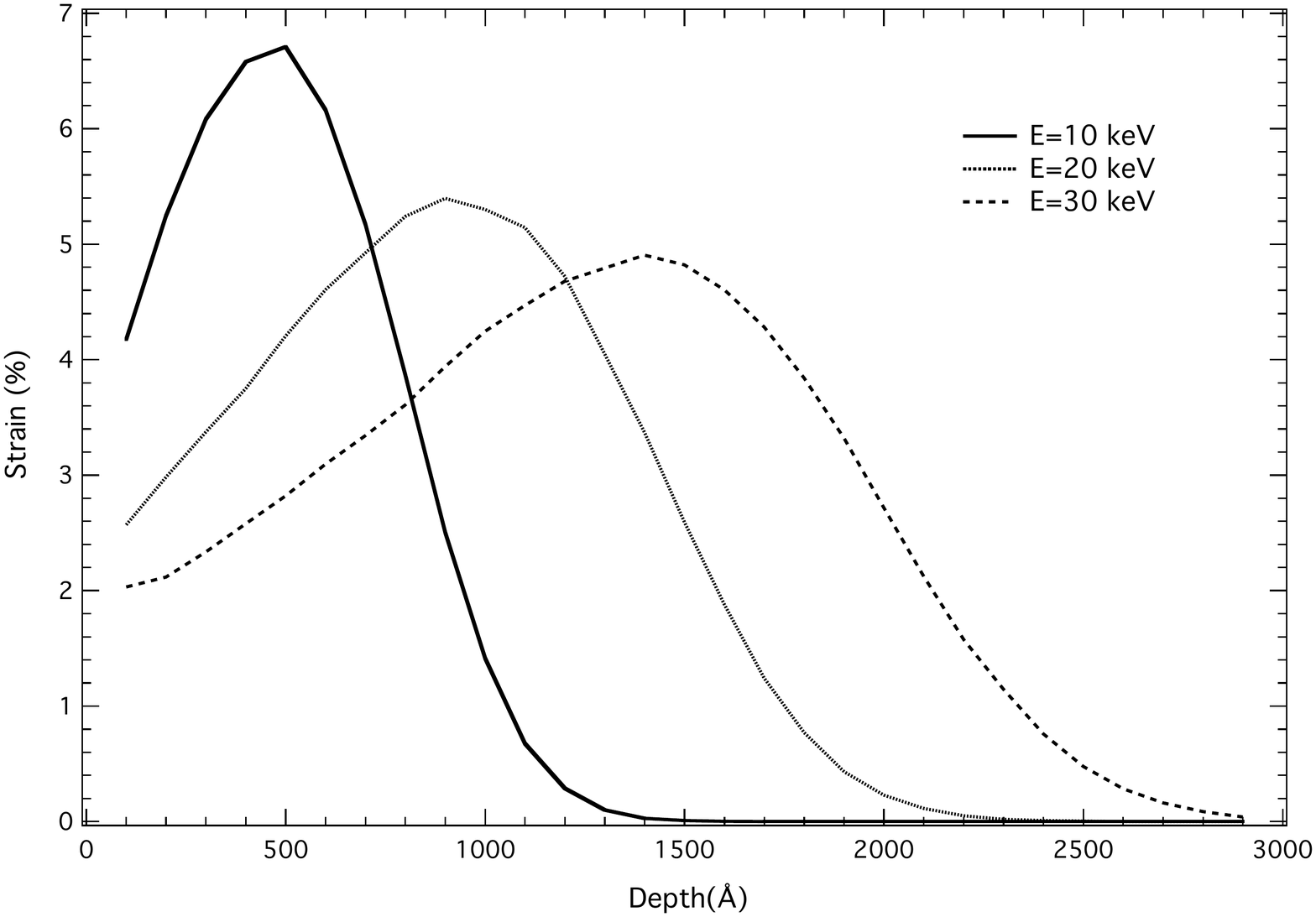}
\caption{Perpendicular strain profiles following the nuclear energy loss distributions, as used in the simulation of the X-ray diffraction measurements (see fig.2), in SrTiO$_3$ irradiated at different He$^{{\rm +}}$ ion energies as indicated and an ion fluence of 10$^{16}$  He$^{{\rm +}}$ ions/cm$^2$.}
\label{Fig4}
 \end{figure}

 The strain in the irradiated area has been studied using X-ray diffraction for both (001) and (110) oriented crystals. For the (001) orientation, the perpendicular lattice parameter was probed around the (002) Bragg peak in  $\theta$-2$\theta$ geometry, while the in-plane lattice parameter was probed with a measurement around a (202) Bragg reflection. For the latter, our measurements showed almost no in-plane strain, less than 0.3\% for all the measured samples irradiated between 10$^{15}$ He$^{{\rm +}}$ ions/cm$^2$ and 10$^{16}$ He$^{{\rm +}}$ ions/cm$^2$. However, the perpendicular strain showed a strong increase after helium ion implantation as can be seen in the measurement presented in Fig.\ref{Fig3} for a sample irradiated with 10$^{16}$ He$^{{\rm +}}$ ions/cm$^2$.  As compared to the reference measurement on a non-irradiated SrTiO$_3$ sample, one observes intensity on the low angle side of the Bragg peak corresponding to lattice expansion in the direction perpendicular to the sample surface. As was shown by Sousbie et al.\cite{Sousbie_JAP2006}, this result can be described rather simply with a direct inversion procedure. This method attributes the interference fringes in the diffraction profile to equivalent strain states situated at different depths along a gaussian strain profile. One can directly obtain a perpendicular strain profile in the sample with a maximum of 5.15\% and FWHM of 166 nm (gaussian profile). A similar result was obtained with the (110) oriented sample and no dependency with surface orientation was observed, as for example the same maximum strain of 2.4\% was observed in the (001) and (110) oriented samples irradiated at 30 keV and 5$\times$10$^{15}$  He$^{{\rm +}}$ ions/cm$^2$. For the experimental data simulation, a more precise approach consists in calculating the diffraction spectrum directly with a X-ray diffraction simulation program (http://sergey.gmca.aps.anl.gov/gid\_sl.html). This method was used starting from different strain profiles and adjusting the maximum strain value to fit with the lowest angle observed interference fringe. The simulations show that a strain profile following the helium concentration profile does not describe properly the data, while a strain profile following the nuclear energy loss profile allows a good description of the data as can be seen from Fig.\ref{Fig3}. The position of the observed maximum around 2$\theta$=45$^\circ$ is seen to be highly sensitive to the strain profile. We infer from the simulations that the strain observed in strontium titanate after helium ion implantation in a range from 5$\times$10$^{15}$  to 10$^{16}$ He$^{{\rm +}}$ ions/cm$^2$ is mainly due to Frenkel pair formation along the ion beam, as was already observed in SiC by Leclerc et al.\cite{Leclerc_JAP2005}. A stress model for thin films including the effect of volume distortions induced by point defects created under ion irradiation was proposed by Debelle et al.\cite{Debelle_APL2004}. Within the framework of this model, the in-plane lattice parameter is fixed by the substrate, giving only one direction of expansion of the layer under irradiation. The observed maximum perpendicular strain was observed to follow a linear dependency  with ion fluence from 0.5\% at 10$^{15}$ He$^{{\rm +}}$ ions/cm$^2$ up to 5\% at 10$^{16}$ He$^{{\rm +}}$ ions/cm$^2$ for an irradiation at 30 keV. This result supports the idea that the strain is proportional to the deposited energy in that fluence regime. For higher fluences (above 5$\times$10$^{16}$  He$^{{\rm +}}$ ions/cm$^2$) and for both surface orientations, the X-ray diffraction data indicates surface amorphization after implantation at room temperature, without any observed interference fringes. The irradiation energy dependency of the perpendicular strain has been probed between 10 keV and 30 keV at a fluence of 10$^{16}$  He$^{{\rm +}}$ ions/cm$^2$. The strain profiles used to simulate the X-ray measurements are shown in Fig.\ref{Fig4}.  At 30 keV, the maximum strain is nearly constant in a band of about 40 nm thickness centered at 140 nm. We also observe a decreasing maximum strain as a function of energy. Indeed,  the maximum strain was about 6.7\% at 10 keV, 5.4\% at 20 keV and 4.9\% at 30 keV. This result can be understood in term of the deposited ion energy profile: as obtained from the SRIM simulation program, the ion energy is deposited in a narrower band at lower energy, giving rise to a higher defect concentration and higher strain in that located area.
 
\subsection{TRANSMISSION ELECTRON MICROSCOPY}

\begin{figure}
\includegraphics[width=4in]{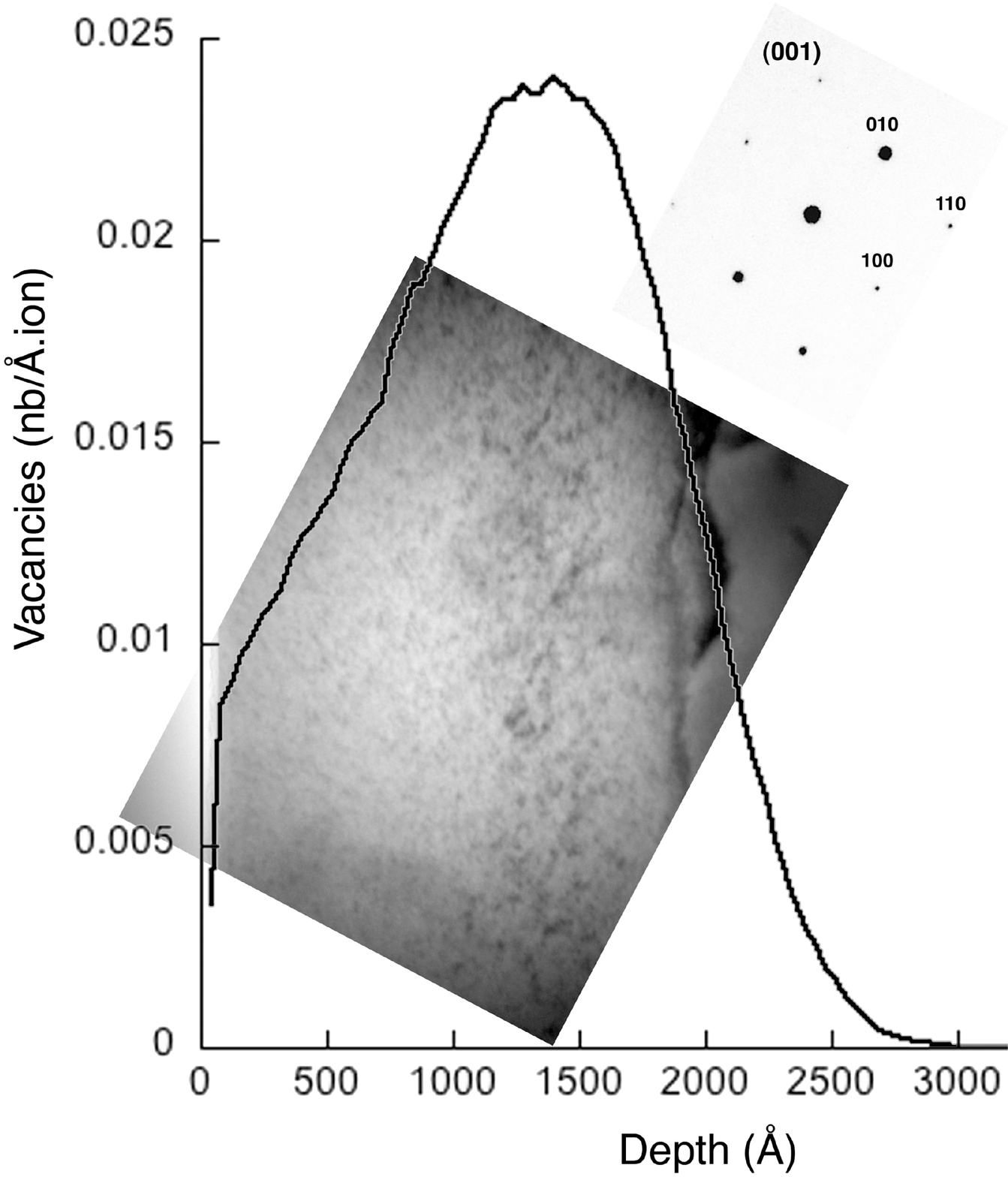}
\caption{Oxygen vacancy concentration profile obtained from SRIM simulation superposed with a TEM bright field image of a (110) oriented SrTiO$_3$ irradiated with 10$^{16}$ He$^{{\rm +}}$ ions/cm$^2$ at 30 keV. The cross-section is observed at a magnification of 390k. The insert shows a selected area diffraction pattern taken in the near-surface band.}
\label{Fig2}
 \end{figure}

In order to illustrate the near-surface sample modification related to helium ion implantation, we show in Fig.\ref{Fig2}, a TEM image of a (110) oriented SrTiO$_3$ sample irradiated with 10$^{16}$ He$^{{\rm +}}$ ions/cm$^2$ at 30 keV. This image clearly shows a band related to the irradiated area with a thickness of 175 nm, while defects are mostly seen in a zone corresponding to the maximum nuclear energy loss. For comparison, the oxygen vacancy profile as obtained from SRIM simulation is superposed, clearly showing a correspondence with the ion induced defects in the irradiated area, peaking around 140 nm. The maximum helium concentration peaks at the end of the observed band (175 nm), indicating that the helium interstitial atoms instead induce strain relaxation as it was observed in SiC\cite{Leclerc_JAP2005,Leclerc_APL2008}. As can be seen in Fig.\ref{Fig2} the strain relaxation between the implanted band and the substrate leads to the formation of extended defects at a depth of about 200 nm. On another hand, the diffraction pattern from the irradiated zone (shown in the insert), probed with an area selection diaphragm, evidences that the sample is still monocrystalline after ion irradiation. The irradiated sample does not show any evidence of amorphization which would give for example diffuse scattering around the observed diffraction spots. Also, no evidence for microcavities has been observed and vacancy like defects might be stabilised by the implanted He atoms\cite{Schut}. A rough measurement of the distance between different Bragg peaks in the irradiated and non-irradiated areas gives an estimation for the mean strain in the observed area of about 1.5\%. This value, as expected, is lower than the one obtained from X-ray diffraction, and indicates consistency between the two measurements. 

\section{CONCLUSION}
We have shown here the possibility, using helium ion irradiation, to modify in a controlled manner the perpendicular lattice parameter of SrTiO$_3$ in a variable thickness layer close to the surface. This near-surface strain distribution has been attributed predominantly to the creation of Frenkel pairs in the studied fluence range. As compared to previous work based on the modification of SrTiO$_3$ using argon ion irradiation at low-energy (300 eV)\cite{Kan_NatMat2005}, our method does not amorphize the sample surface. It also introduces a buried and tunable perpendicular strain  distribution, while argon ion irradiation have been shown to induce lattice parameter expansion only in a thin surface layer\cite{Kan_Jpn2007}. The strain gradients introduced by helium ion irradiation could be used to generate polarization in SrTiO$_3$\cite{Zubko_2007}, though we did not observe any second harmonic generation (SHG) signal at room temperature. The possibility to tune the oxygen vacancy concentration at a given depth may open the way for a controlled introduction of charge carriers in thin buried layers, and thus to induce new properties in oxide materials\cite{Reyren_Science2007}. Another possibility offered by the method will be the use of the local swelling associated with the perpendicular lattice parameter expansion for surface nano-patterning\cite{Albrecht_SS2003,Lindner_NIMB2009,Lang_IJN2009}.

\begin{acknowledgments}
We thank J. Moeyaert for the help with the implantation, and A. Debelle for discussion about the X-ray analysis. The authors acknowledge I. Genois and S. Collin for TEM sample preparation and V. Pillard for contributing in the X-ray measurements.
S. Autier-Laurent acknowledges the C'Nano Ile de France for financial support.
\end{acknowledgments}


\newpage

\begin{thebibliography}{99}
 
\bibitem{Prellier_JAP2001}Prellier W.,Simon C.,Mercey B.,Hervieu M., Haghiri-Gosnet A. M., Saurel D., Lecoeur P., and Raveau B., J. Appl. Phys. {\bf 89}(2001) 6612.
 \bibitem{Haeni}Haeni J. H., Irvin P., Chang W., Uecker R. , Reiche P., Li Y. L., Choudhury S., Tian W., Hawley M. E., Craigo B., Tagantsev A. K., Pan X. Q., Streiffer S. K., Chen L. Q., Kirchoefer S. W., Levy J.  and Schlom D. G., Nature {\bf 430} (2004) 758.
\bibitem{Chaumont_NIMB1981}Chaumont J., Lalu F., Salom\'e M., Lamoise A.-M. and Bernas H., Nucl. Instrum. Methods Phys. Res. {\bf 189} (1981) 193.
\bibitem{SRIM} Ziegler J., Biersack J. and Littmark U., The Stopping of Ions in Matter, Pergamon New York (1985).
\bibitem{Schut}Schut H., Van Veen A., Labohm F., Fedorov A. V., Neeft E. A. C.  and Konings R. J. M., Nucl. Instrum. Methods Phys. Res. B {\bf 147} (1999) 212.
\bibitem{Gentils}Gentils A., Copie O., Herranz G., Fortuna F., Bibes M., Bouzehouane K., Jacquet \'E., Carr\'et\'ero C., Basleti\'c M., Tafra E., Hamzi\'c A. and Barth\'el\'emy A., Phys. Rev. B {\bf 81}(2010)144109.
\bibitem{Sousbie_JAP2006}Sousbie N., Capello L., Eymery J., Rieutord F.  and Lagahe C., J. Appl. Phys.{\bf 99}(2006)103509.
\bibitem{Leclerc_JAP2005}Leclerc S., Decl\'emy A., Beaufort M. F., Tromas C.  and Barbot J. F., J. Appl. Phys..{\bf 98}(2005)113506.
\bibitem{Debelle_APL2004}Debelle A., Abadias G., Michel A. and Jaouen C., Appl. Phys. Lett..{\bf 84}(2004)5034.
\bibitem{Leclerc_APL2008}Leclerc S., Beaufort M. F., Decl\'emy A.  and Barbot J. F., Appl. Phys. Lett.{\bf 93}(2008)122101.
\bibitem{Kan_NatMat2005}Kan D., Terashima T., Kanda R., Masuno A., Tanaka K., Chu S., Kan H., Ishizumi A., Kanemitsu Y., Shimakawa Y. and Takano M., Nature Mat.{\bf 4}(2005)816.
\bibitem{Kan_Jpn2007}Kan D., Sakata O., Kimura S., Takano M. and Shimakawa Y., Jpn. J. Appl. Phys.{\bf 46}(2007)L471.
\bibitem{Zubko_2007}Zubko P.., Catalan G., Buckley A., Welche P.R.L. and Scott J.F., Phys. Rev. Lett.{\bf 99}(2007)167601.
\bibitem{Reyren_Science2007}Reyren N., Thiel S., Caviglia A. D., Fitting Kourkoutis L., Hammerl G., Richter C., Schneider C. W., Kopp T., Ruetschi A.-S., Jaccard D., Gabay M., Muller D. A., Triscone J.-M.  and Mannhart J., Science.{\bf 317}(2007)1196.
\bibitem{Albrecht_SS2003}Albrecht J., Leonhardt S., Spolenak R., Taffner U., Habermeier H. U. and Schutz G., Surf. Sci.{\bf 547}(2003)L847.
\bibitem{Lindner_NIMB2009}Lindner J. K. N., Seider C., Fischer F., Weinl M. and Stritzker B., Nucl. Instrum. Methods Phys. Res. B.{\bf 267}(2009)1394.
\bibitem{Lang_IJN2009}Lang W., Richter H., Marksteiner M., Siraj K., Bodea M. A., Pedarning J. D., Grigoropoulos C., Bauerle D., Hasenfuss C., Palmetshofer L., Kolarova R.  and Bauer P., Int. J. Nanotech. {\bf 6}(2009)704.

\end{thebibliography}
\end{document}